# Discovery of Intrinsic Quantum Anomalous Hall Effect in Organic Mn-DCA Lattice

Ya-ping Wang,[1,2] Wei-xiao Ji,[1] Chang-wen Zhang*,[1] Ping Li,[1] Pei-ji Wang [1], Biao Kong,[2] Sheng-shi Li,[3] and Shi-shen Yan,[3] Kang Liang [4]

[1] *School of Physics and Technology, University of Jinan, Jinan, Shandong, 250022, PR China*

[2] *Department of Chemistry, Shanghai Key Lab of Molecular Catalysis and Innovative Materials, iChEM, Fudan University, Shanghai 200433, PR China*

[3] *School of Physics, State Key laboratory of Crystal Materials, Shandong University, Jinan, Shandong 250100, PR China*

[4] *School of Chemical Engineering, The University of New South Wales, Sydney, NSW 2052, Australia*

**Abstract:** The quantum anomalous Hall (QAH) phase is a novel topological state of matter characterized by a nonzero quantized Hall conductivity without an external magnetic field. The realizations of QAH effect, however, are experimentally challengeable. Based on *ab initio* calculations, here we propose an intrinsic QAH phase in Mn−dicyanoanthracene (DCA) *Kagome* lattice. The nontrivial topology in *Kagome* bands are confirmed by the nonzero chern number, quantized Hall conductivity, and gapless chiral edge states of Mn-DCA lattice. A tight-binding (TB) model is further constructed to clarify the origin of QAH effect. Furthermore, its Curie temperature, estimated to be ~ 253 K using Monte-Carlo simulation, is comparable with room temperature and higher than most of two-dimensional ferromagnetic thin films. Our findings present a reliable material platform for the observation of QAH effect in covalent-organic frameworks.

**Keywords:** Quantum anomalous Hall effect, Organometallic framework, Dirac band, Spin-orbit coupling

Corresponding author: E-mail: ss_zhangchw@ujn.edu.cn (C. W. Zhang)

Two-dimensional (2D) topological insulators (TIs), also known as quantum spin Hall (QSH) insulators, are an exotic quantum state of matter with dissipationless chiral edge states protected by the electron band topology, where spin-orbit coupling (SOC) plays a role analogous to the external magnetic field in quantum Hall systems.[1-6] When a ferromagnetic (FM) exchange interaction is introduced, the time-reversal symmetry (TRS) is disrupted and the chemical potentials of two metallic channels with opposite spin polarizations can be unbalanced, leading to the so called quantum anomalous Hall (QAH) effect. [7-10] The spin-polarized edge electron channel in QAH insulators is of crucial importance due to its incredibly precise quantization, robustness against defects, disorder and surface contamination over hundreds of micrometers [11], which are highly promising for the design of new quantum devices such as the chiral interconnect. [9]

The QAH model is first proposed by Haldane in 2D honeycomb lattice, in which the SOC and FM ordering realize the intrinsic QAH effect.[12] Subsequently, the main proposed materials for the QAH phase are based on graphene and silicene doped with external metal atoms.[13–16] However, these 2D stoichiometric magnetic honeycomb lattices exist rarely in nature and are difficult to manipulate due to their structural complexity. The *Kagome* lattice in metal-organic framework (MOF), which shares similar geometric structure with the honeycomb lattice, has attracted much interest in solution-processed organic or inorganic devices. [17-21] Remarkably, some of these 2D MOFs has been predicted to be topological phases [22-26], also named as organic topological insulators (OTIs). Compared to the inorganic materials,[13-16] these 2D MOFs have potentially the advantages of low cost, easy fabrications, and mechanical flexibility. Unfortunately, only limited 2D OTIs proposed in a *Kagome* lattice is intrinsic and these materials still need additional holes or electrons to move the Fermi level to inside the bulk gap induced by SOC. [22–27] For example, in a 2D HTT-Pt lattice, the doping electrons at a level of $8.28 \times 10^{13}$ cm$^{-2}$ is required to move the Fermi level to topologically nontrivial gap, which is very challenging experimentally. [26] The experimental realization of these exciting QAH states is still in infancy.

In this letter, based on *ab initio* density functional theory (DFT) and tight-binding (TB) model, we investigate the electronic and topological properties of a new *Kagome* lattice, Mn-dicyanoanthracene (DCA), in an existing synthesized MOFs. [28,29] Considering that the half-filled Mn atom has the $3d^54s^2$ valence-electron configuration, it is expected that the Mn-DCA lattice would be strongly spin-polarized to realize a QAH state, similar to the case of HTT-Pt lattice.[26] Indeed, the calculated Chern numbers, quantified Hall conductivity, and gapless chiral edge states within the SOC gaps, confirm the nontrivial topology of the Mn-DCA lattice. Remarkably, the QAH phase is intrinsic with the Fermi level located in the nontrivial *Kagome* gap, without tuning the holes or electrons of Mn-DCA lattice. These findings demonstrate that the Mn-DCA lattice can expedite experimental advancement on QAH phase and its practical applications in spintronics devices.

All the *ab initio* calculations are carried out based on projector-augmented wave method[30] with exchange-correlation functional in the Perdew-Berke-Ernzerhof (PBE) form within the generalized-gradient approximation,[31] as implemented in VASP package.[32,33] Fully optimized lattice constant $a$ = 20.66 Å and a strong coordination bond of a Mn-N bond length of 1.94 Å in the force smaller than 0.01 eV/Å are obtained, which are similar to the experimental results on Metal-DCA lattices.[28, 29] The cutoff energy for wave function expansion is set to 500 eV, and 15×15×1 k meshes are used for the slab self-consistent calculations. SOC is included by a second variational procedure on a fully self-consistent basis.

Figure 1(a) shows the geometric lattice of the 2D Mn-DCA lattice, where Mn atoms form a honeycomb lattice with a threefold rotational symmetry by sharing the DCA bridges with three neighbors. Thus, the half-filled 3*d* shell of Mn is strongly spin polarized with a magnetic moment of 4.0 $\mu_B$/Mn as a result of large Hund's rule coupling. The calculated total energy of the FM ground state is 3.67 and 5.86 eV/f.u. lower than the antiferromagnetic (AFM) and nonmagnetic (NM) states, respectively, suggesting that the Mn-DCA lattice favors an FM exchange coupling between Mn atoms. Furthermore, the magnetic anisotropy calculation demonstrates that the FM ground state exhibits the out-of-plane spin orientation, which is 55 meV lower than the in-plane spin orientation. By performing Monte-Carlo simulations, we can obtain the FM Curie temperature of Mn-DCA lattice to be as high as 253 K (Fig. S2). This real-space property is confirmed by the analysis of spin density in Fig. S1, and could be explained by the interacting localized moments on Mn rather than by the itinerant mechanism of ferromagnetism.

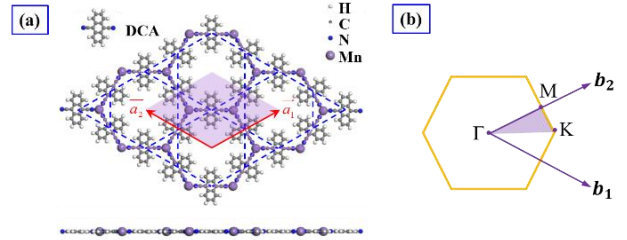

**Fig. 1** (a) Top and side view of the optimized 2D Mn-DCA lattice structure with lattice vectors and in the *xy* plane. Rhombus shows the unit cell. (b) the first Brillouin zone of the structure, with reciprocal lattice vectors and high symmetry points Γ, M, and K.

The calculated band structures of Mn-DCA lattice without SOC are shown in Fig. 2(a), where the red solid lines display the spin-up channel, while the blue dashed lines display the spin-down one. There is a typical *Kagome* band around the Fermi level, consisting of one completely flat band above two Dirac bands remaining at K point, see Fig. 2(c). Different from the case of a nonmagnetic Cu-DCA lattice,[34] the spin degeneracy of the *Kagome* band is removed with a split of about 0.40 eV, due to the internal magnetization from Mn atoms. More interestingly, the Fermi level crosses a Dirac band of spin-up channel but resides at the band gap (0.76 eV) of spin-down channel, displaying the half-metal characteristic. Further analysis of the orbital projected electronic states for the Mn-DCA lattice clearly demonstrates that the *Kagome* bands near the Fermi energy mostly come from DCA molecules, along with less of the Mn-*3d* states.

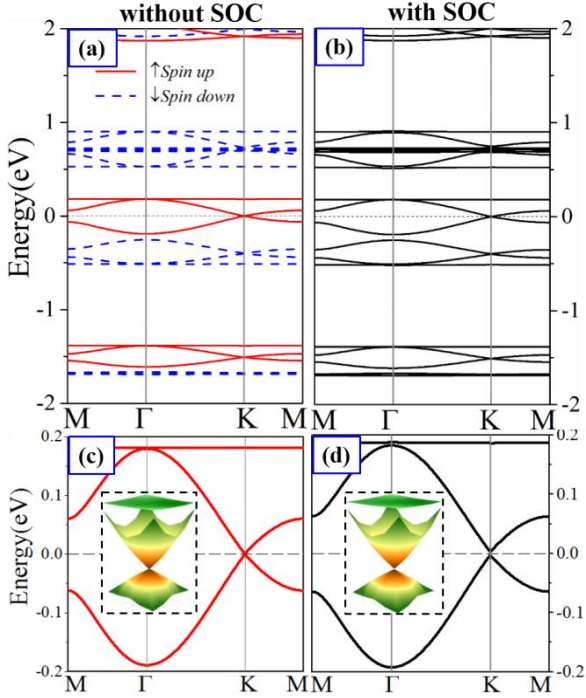

**Fig. 2** (a), (b) Band structures of the Mn-DCA lattice without and with SOC, respectively. Red solid lines and blue dashed lines denote spin-up and spin-down bands. (c), (d) Magnification of (a) and (b) around the Fermi level, respectively, while the inserts are 3D-bands of *Kagome* band near Feimi level.

When the SOC is included, a band gap of 2.3 meV opens at the Dirac point with the Fermi level crossing exactly inside the bulk SOC gap, as shown in Figs. 2(b) and (d). Such a band gap opening at the *K* point suggests a topologically nontrivial feature. To confirm this nontrivial topology of Mn-DCA lattice, the Berry curvature $\Omega(k)$ from the Kubo formula [35,36] is calculated by

$$\Omega(k) = \sum_n f_n \Omega_n(k)$$

$$\Omega_n(k) = -2\,\mathrm{Im} \sum_{m \neq n} \frac{\langle \psi_{nk}|v_x|\psi_{mk}\rangle \langle \psi_{mk}|v_y|\psi_{nk}\rangle \hbar^2}{(E_m - E_n)^2}$$

where the summation is over all of the occupied states, $E_n$ is the eigenvalue of the Bloch function $|\psi_{nk}\rangle$, $f_n$ is the Fermi-Dirac distribution function, and $v_x$ and $v_y$ are the velocity operators. Figure 3(a) shows the Berry curvature for the whole valence bands along the high-symmetry directions in momentum space. One can see that the nonzero Berry curvatures mainly localize around K and K′ points with the same sign. By integrating the $\Omega(k)$ over the first Brillouin zone (BZ), the Chern number (*C*),

$$C = \frac{1}{2\pi} \sum_n \int_{BZ} d^2k\, \Omega_n ,$$

is obtained to be -1, with each Dirac cone (K and K′) contributing -0.5. Thus, the anomalous Hall conductivity $(\sigma_{xy})$, $\sigma_{xy} = \frac{e^2}{h} C$, shows a quantized charge Hall plateau of at a value of $\frac{e^2}{h}$ located in the insulating gap of the spin-up Dirac cone. Such a nonvanishing Chern number and quantized Hall conductivity indicate the QAH phase in the Mn-DCA lattice.

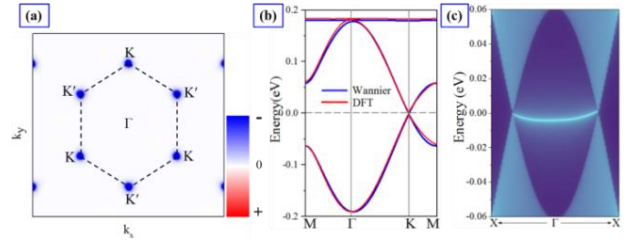

**Fig. 3** (a) The Berry curvature with SOC in the momentum space. The red-white-blue color gives distribution of Berry curvature from positive to negative value in arbitrary unit, and the black dash lines show the first BZ. (b) DFT and MLWFs fitted band structures Mn-DCA. (c) Electronic structure of helical edge states of Mn-DCA, shows the total density of states.

The existence of topologically protected chiral edge states is one of the most important signatures of the QAH effect. To further reveal the nontrivial topological nature of Mn-DCA lattice, we constructed the Green's functions [37] for the semi-infinite boundary based on the maximally localized Wannier function method, [38,39] and obtained the local density of state (LDOS) of the edge states, as shown in Fig. 3(c). Obviously, the nontrivial edge states connecting the valence and conduction bands cross the insulating gap of the spin-up Dirac cone, is consistent with the Chern number $C = -1$. The spin-polarized Dirac-fermion mediated topological characters suggest that the Mn-DCA lattice proposed here is an intrinsic magnetic TI, holding the potential the QAH effect in spintronics.

As shown in Figs. 2(b) and (d), the spin-up and spin-down bands are no longer separable when the SOC is included, but the z-component of the spin is a conserved quantum number since the magnetic order of Mn-DCA lattice is perpendicular to Mn-DCA plane. Thus the Hamiltonian conserves the z-component of the spin, and therefore we can divide the one-particle Hilbert space H into $H_\uparrow \oplus H_\downarrow$ by the eigenvalue of $\sigma^z$. To reproduce the *Kagome* band in Mn-DCA lattice, we present the space $H_\uparrow$ because there are only spin-up bands near the Fermi energy. The single-orbital TB model on Mn-DCA lattice can be expressed as

$$H = \begin{pmatrix} E_0 & 2t\cos k_1 & 2t\cos k_2 \\ 2t\cos k_1 & E_0 & 2t\cos k_3 \\ 2t\cos k_2 & 2t\cos k_3 & E_0 \end{pmatrix}$$

$$\pm i2\lambda \begin{pmatrix} 0 & \cos k_1 & -\cos k_2 \\ -\cos k_1 & 0 & \cos k_3 \\ \cos k_2 & -\cos k_3 & 0 \end{pmatrix}$$

where $k_1 = (a/2)\hat{x}$, $k_2 = (a/2)[(\hat{x}+\sqrt{3}\hat{y})/2]$, $k_3 = (\frac{a}{2})[\frac{-\hat{x}+\sqrt{3}\hat{y}}{2}]$, and $a$ is the lattice constant. $E_0$ is the on-site energy, $t$ is the nearest-neighbor hopping parameter, $\lambda$ is the nearest-neighbor intrinsic SOC, and ± refers to the spin-up/spin-down bands. Figure 3(b) presents the calculated band structure with the inclusion of SOC. Remarkably, the linear Dirac bands along the high-symmetry directions crosses at the K point with a small gap of 2.3 meV, which is consistent with the obtained DFT results. The corresponding fitting parameters for the Mn-DCA lattice are $E_0 = 0.0605$ eV, $t = -0.0605$ eV, and $\lambda = 0.0004$ meV. These demonstrate that the intrinsic SOC in Mn-DCA lattice is responsible for gap opening at the Dirac bands, which is of importance to realize the QAH effect.

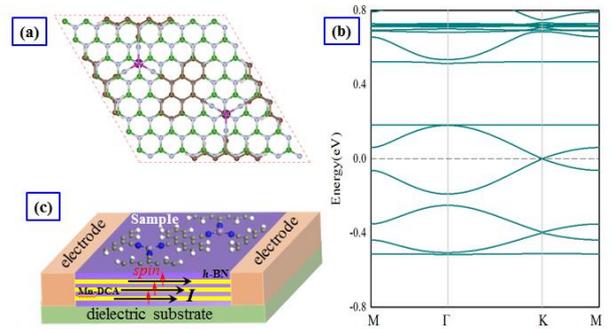

**Fig. 4** (a) Top view of the epitaxial growth of the Mn-DCA lattice on 8×8 BN substrate. (b) The corresponding energy band structure with SOC. (c) Schematic model for proposed Mn-DCA/BN heterostructure for quantum state measurement. Vertical arrows show the spin orientation of electrons in the edge states and horizontal arrows show their transport directions.

We further check the correlation effect in 3$d$ electrons of transition-metal Mn on electronic properties. The band structures calculated within GGA+$U$ ($U$ = 3.0 eV is the parameter for onsite Coulomb interaction) are similar with the initial conditions. The correlation effect is negligible and the ground states of Mn-DCA lattice are still spin-polarized with a magnetic moment 4.0 $\mu_B$, and thus the results discussed above are robust. [40]

Finally, one critical point is whether the QAH effect of the Mn-DCA lattice can remain on a substrate, since the substrates are inevitable in device applications. To test out this possibility, we place the Mn-DCA lattice on top of 8×8 BN substrate to form a Mn-DCA/BN heterostructure, as shown in Fig. 4(a). After structural optimization, the distance of Mn-DCA and BN layers is 3.51 Å with a binding energy of -58 meV per unit cell, suggesting a typical vdW structure. In this case, the main features of QAH effect in free-standing Mn-DCA lattice remain intact. Figure 4(b) presents the calculated band structure with SOC. As expected, in these weakly coupled systems, there is still a SOC gap at the Dirac point around the Fermi level, and the states around the Fermi level are dominantly contributed by the *Kagome* band. Considering the 2D wide-gap BN sheet electrically insulate adjacent QSH layers of Mn-DCA lattice, protecting parallel helical edge channels from being gapped by interlayer hybridization, the predicted Mn-DCA/BN heterostructure can parametrically increase the number of edge transport channels to support the dissipationless charge/spin transport in the topological states. These results demonstrate the feasibility of constructing quantum devices by Mn-DCA/BN heterostructure, as illustrated in Fig. 4(c).

In conclusion, we demonstrate the possibility of realizing the intrinsic QAH effect in 2D *Kagome* lattice, and predict that the Mn-DCA lattice is a good candidate. The Curie temperature estimated from Monte Carlo simulations within the Ising model is about 253 K. Also, the nontrivial properties in *Kagome* bands are confirmed by a nonzero chern number, quantized Hall conductivity, and gapless chiral edge state. A TB model is constructed to explain the origin of nontrivial topology. Such 2D MOF materials are much easier to synthesize and much more homogeneous than inorganic materials, therefore enabling the Mn-DCA lattice as a more promising platform for realizing low-dissipation quantum and spintronics devices.

---

**Supplementary Material:** See supplementary material for Discovery of Intrinsic Quantum Anomalous Hall Effect in Organic Mn-DCA Lattice.

**Acknowledgments:** This work was supported by the National Natural Science Foundation of China (Grant NFS. 11434006 and 61571210).